\documentclass[prl,twocolumn]{revtex4}

\usepackage{graphicx}
\usepackage{amsmath} 
\usepackage{amsfonts} 
\usepackage{amssymb}
\usepackage{bm}
\usepackage{textcomp}
\usepackage{float}
\usepackage[colorlinks]{hyperref}
\usepackage{epstopdf}

\newcommand{\RR}{\right}
\newcommand{\LL}{\left}
\newcommand{\m}{\mathrm}

\begin{document}
\title{Strong gate coupling of high-\emph{Q} nanomechanical resonators}

\author{Jaakko Sulkko$^1$}
\author{Mika A. Sillanp\"a\"a$^1$}
\author{Pasi H\"akkinen$^1$}
\author{Lorenz Lechner$^1$}
\author{Meri Helle$^1$}
\author{Andrew Fefferman$^2$}
\author{Jeevak Parpia$^3$}
\author{Pertti J. Hakonen$^1$}

\affiliation{$^1$Low Temperature Laboratory, Aalto University, P.O. Box 15100, FI-00076 AALTO, Finland\\
$^2$University of California, Berkeley, California 94720, US \\
$^3$Department of Physics and LASSP, 608 Clark Hall, Cornell University, Ithaca NY 14853-2501 USA}



\begin{abstract}
The detection of mechanical vibrations near the quantum limit is a formidable challenge since the displacement becomes vanishingly small when the number of phonon quanta tends towards zero. An interesting setup for on-chip nanomechanical resonators is that of coupling them to electrical microwave cavities for detection and manipulation. Here we show how to achieve a large cavity coupling energy of up to $(2 \pi) \, 1$ MHz/nm for metallic beam resonators at tens of MHz. We used focused ion beam (FIB) cutting to produce uniform slits down to 10 nm, separating patterned resonators from their gate electrodes, in suspended aluminum films. We measured the thermomechanical vibrations down to a temperature of 25 mK, and we obtained a low number of about twenty phonons at the equilibrium bath temperature. The mechanical properties of Al were excellent after FIB cutting and we recorded a quality factor of $Q \sim 3 \times 10^5$ for a 67 MHz resonator at a temperature of 25 mK. Between 0.2K and 2K we find that the dissipation is linearly proportional to the temperature. \\
 \\
\textbf{Keywords \\}
nanomechanics, NEMS, quantum limit, detection, dissipation
\end{abstract}
\maketitle

The measurement of small-amplitude vibrations in mechanical systems is becoming an increasingly interesting problem~\cite{Clelandbook,Ekinci2005}. From the point of view of basic science, the study of mechanical systems close to the quantum limit has attracted a lot of interest recently~\cite{SchwabQ,LaHaye2004}. The endeavor towards the ground state of the harmonic phonon oscillations has been going on in various physical systems such as in optomechanics~\cite{Karrei,Gigan,Kippenberg}, or in electrically coupled beam resonators which have been measured using single-electron transistors~\cite{Knobel,LaHaye2004}, or lately, with on-chip microwave cavities~\cite{Truitt2007,Regal2008,Sillanpaa2009,Schwab10}.

The quantum challenge is posed by several issues, including the relatively low frequency ($f_0 \sim 10$ MHz), of the lowest modes in suspended beams. The quantum limit implies stringent requirements on temperature, since $h f_0$ needs to be small in comparison to $k_B T$. On the other hand, at higher frequencies, the coupling to measuring systems diminishes rapidly. Third, the zero-point vibration amplitudes $x_{\m{ZP}} = \sqrt{\hbar /2 m \omega_0}$, where $m$ is the effective mass and $\omega_0 = 2 \pi f_0$ is the angular frequency, are vanishingly small even at the atomic scale. Very recently, O'Connell \emph{et al.}~\cite{ClelandMartinis} demonstrated a piezoelectric mechanical mode at the quantum ground state by using a coupling to a superconducting qubit. However, bringing a purely mechanical mode to the quantum limit remains an ongoing quest, with the goal becoming a reality probably in the near future.

\begin{figure*}
 \includegraphics[width=16cm]{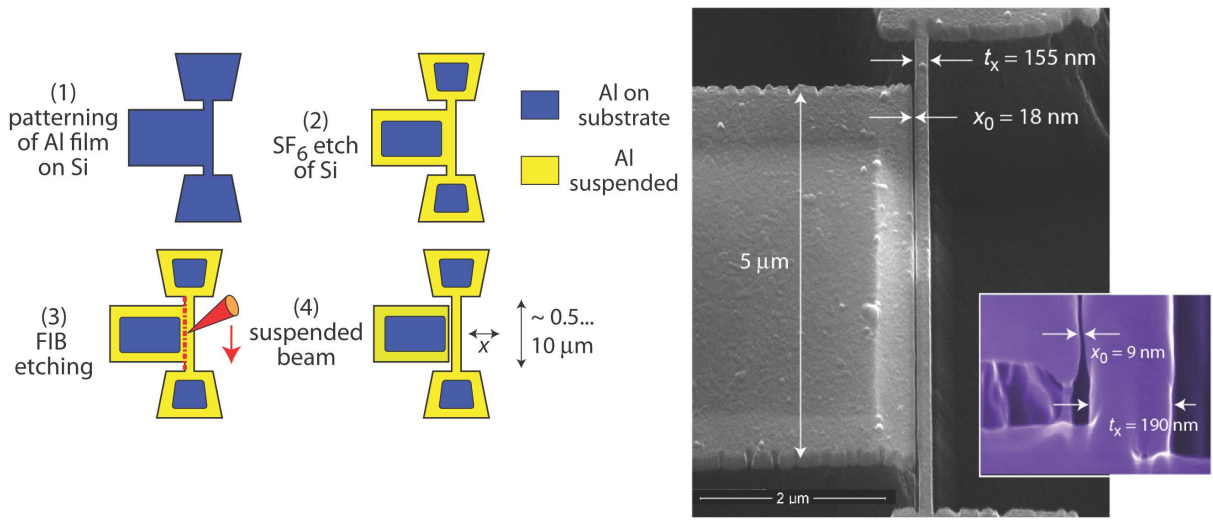}
  \caption{The fabrication of strongly capacitive coupled nanoresonators, (1)-(4), on top of a high-resistivity silicon wafer. E-beam lithography was used to pattern the metallization. Before Focused Ion Beam (FIB) milling, the e-beam evaporated aluminum T-structure was suspended by an optimized SF$_6$ Reactive Ion Etch. The single-pass line cut, with current 10~pA, milling depth $\sim$~100~nm, and $\varnothing\sim$~5~nm at 30~kV, gives vacuum gaps down to 10~nm without any noticeable bending of the beam. On right, SEM image of a beam-gate structure (sample C). The detail shows a short beam with a $< 10$ nm gap, and typical rounded cross section seen after milling. }\label{fig:NEMS}
\end{figure*}

Micromechanical resonators are also used in applications as detectors. The best devices take advantage of the trend to smaller size and higher frequencies, and will soon approach sensing at the atomic mass unit level~\cite{massdetect,massdetectB}. They could also be operated as sensors of position, force, or high-frequency electromagnetic fields.

For conductive resonators, capacitive coupling to an electrical measuring apparatus is useful for readout. In contrast to magnetomotive or optical detection, one can obtain very low electrical losses. An impediment to capacitive measurement is the reduction of coupling with increasing frequency that accompanies reduction in size. The coupling via a capacitance $C_g$ is proportional to $d C_g/dx$, which for typical geometries depends on the vacuum gap $x_0$ between the vibrating beam and the gate electrode as $(x_0)^{-2}$, and hence small gaps are crucial. Using traditional e-beam fabrication, it is hard to obtain $x_0$ below 100 nm. Here we show that focused ion beam (FIB) milling yields gaps down to $\sim 10$ nm for resonator beams up to 10 micron in length and $100-200$ nm in thickness. An improvement in coupling by two orders of magnitude can thus be achieved using FIB-techniques. We also find that mechanical dissipation is reduced by the FIB milling, and we observe quality factors approaching half a million at dilution refrigerator temperatures.

\begin{figure*}
 \includegraphics[width=17cm]{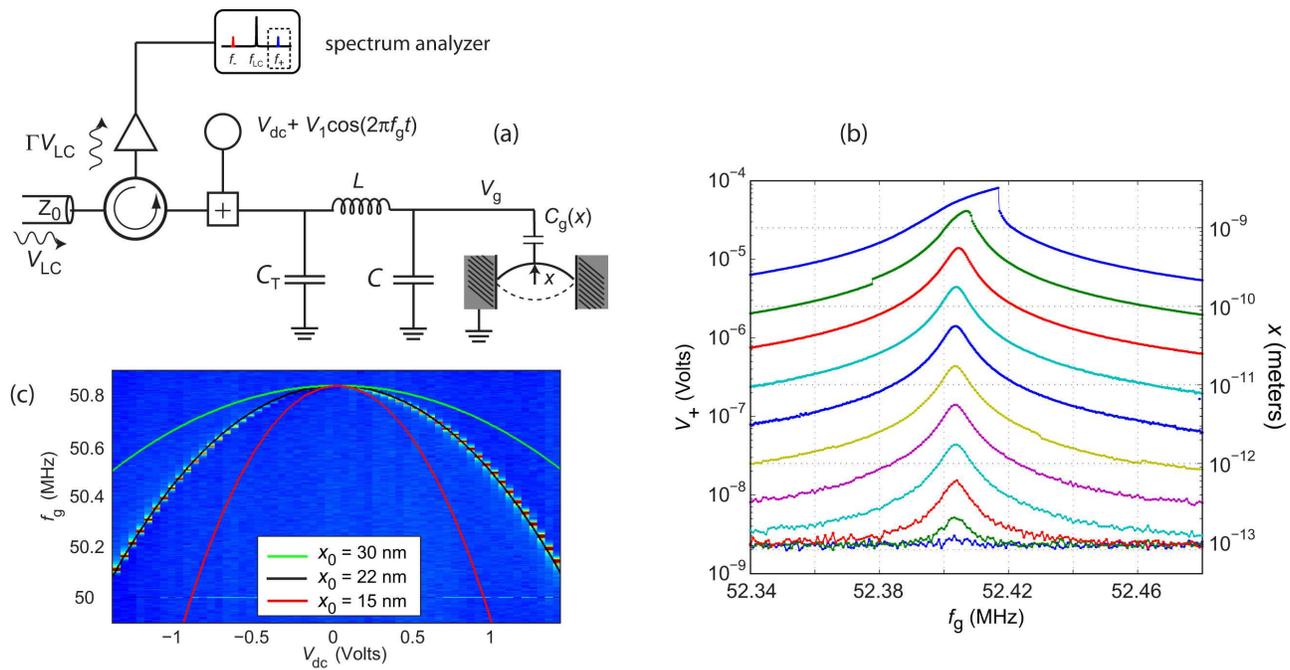}
  \caption{(a) Reflection measurement scheme for the detection of nanomechanical vibrations coupled capacitively to the electrical resonance frequency of a cavity (lumped element $LC$ circuit in the figure), $f_{\m{LC}} = f_{\m{LC}}(1 + g x)$. The vibrations of
the beam at the mechanical frequency $f_0 \ll f_{\m{LC}} \simeq (L C)^{-1/2}/2\pi$ change the total capacitance. The resulting sideband voltage is measured by a spectrum analyzer. The inductive coupling of the cavity allows for introducing a dc voltage to the gate; (b) electrically actuated response of a 5.4-$\mu$m-long nanomechanical resonator (sample D) as a function of mechanical drive frequency $f_g$ measured at $T = 4.2$ K, with voltages $V_{\m{dc}} = 0.2$ V and $V_{\m{LC}} \simeq 1$ mV, while the drive for the mechanical mode $V_1$ is increased from bottom (60 nV) to top (6 mV) at 10 dB intervals; (c) resonance measured as a function of dc bias $V_{\m{dc}}$ (sample C). The solid curves display the calculated dependence at gap values $x_0$ indicated in the figure.}\label{Readout}
\end{figure*}

Our system, as shown in Fig.~\ref{Readout}, is a flexural beam coupled capacitively to an $LC$ cavity resonator detuned substantially from the beam's frequency, {\it i.e.},  $f_{\m{LC}} \gg f_0$. The beam displacement $x$ pulls the cavity frequency by the amount $g x$, where the coupling energy is $g = \LL( \omega_{\m{LC}}/2C \RR) \LL( \partial C_g/\partial x\RR)$, and $\omega_{\m{LC}} = 2 \pi f_{\m{LC}}$. In the quantum language, the Hamiltonian of the whole system is $\hat{H} = \hbar \omega_{\m{LC}} \hat{a}^{\dag} \hat{a} +\hbar \omega_{0} \hat{b}^{\dag} \hat{b} + \hbar g \hat{x} \hat{a}^{\dag} \hat{a} - i g_V (\hat{b} + \hat{b}^{\dag})(\hat{a} - \hat{a}^{\dag} )$. Here, $\hat{a}^{\dag}$ and $\hat{a}$ are the creation and annihilation operators of the cavity, and similarly $\hat{b}^{\dag}$ and $\hat{b}$ for the nanoresonator. $\hat{x} = x_{\m{ZP}} \LL( \hat{b}^{\dag} +\hat{b}\RR)  $ is the quantum amplitude of the nanoresonator. The photon-exchange linear coupling with the energy $g_V = x_{\m{ZP}} V_g \sqrt{\frac{\hbar \omega_{\m{LC}}}{2C}} \LL( \partial C_g/\partial x\RR)$ can be tuned up to a very high value with a dc voltage $V_{\m{dc}}$, but is negligible unless the cavity and the nanoresonator are close to resonance.

We have used \cite{Sillanpaa2009} shorter beams, $l \sim 1.5 ... 10 \, \mu$m, than in Refs.~\onlinecite{Regal2008,Schwab10} and we obtain higher frequencies \cite{Li2008}. Therefore, one would obtain low mechanical occupancies already at dilution refrigerator temperatures. The tradeoff, however, is the smaller area for detection due to the shorter beam length when measuring via an $LC$ cavity, as well as a reduction of gain as $\propto f_{\m{LC}}/f_0$. To this end, FIB processing offers a well suited method to make extremely narrow vacuum gaps for relatively short beams.

In our fabrication, we used focused ion beam (FIB) milling to separate the gate and the beam from a prefabricated, T-shaped, suspended structure, Fig.~\ref{fig:NEMS}. The prefabrication followed closely the steps presented in Ref.~\onlinecite{Sillanpaa2009}. A brief anneal at 250 C for 30 min was performed to generate a built-in tensile strain before cutting. In order to avoid contamination, the FIB modification was performed without exposing nearby structures to the Ga-beam. This was achieved by dual beam operation with careful across-beam alignment (Dual Beam Microscope: Helios Nanolab, FEI Ltd.). The line cut was performed at 30 keV in single-pass mode, instead of gradually scraping the surface, in order to exploit differential sputtering effects which allow smaller gap dimensions and significantly reduce Al redeposition onto the structure. Vacuum gaps of 50-10~nm were obtained by an appropriate choice of the ion beam diameter (min $\varnothing\sim$~5~nm). The dimensions of the measured samples are listed in Table~\ref{sampledimensions}.

\begin{table}
\caption{Length $l$, width $t_x$, height $t_y$, and the vacuum gap $x_0$ of the measured nanomechanical resonators. The resonance frequencies $f_0$ were measured at 4.2 K in a vacuum better than $\sim10^{-4}$ mBar. The estimated frequency is given by $f_0 = \frac{t_x}{l^2} \sqrt{\frac{E}{\rho}}\sqrt{ 1 + \varepsilon \frac{ t_x t_y l^2}{4 \pi^2 I}}$, where $E$, $I=\frac{t_y t_x^3}{12}$, and $\rho$ are, respectively, the Young's modulus, the second moment of inertia, and the density. $\varepsilon = \Delta l/l \sim 0.6\%$ denotes the tensile strain. $f_0^{\m{th}}$ and $f_0^{\m{exp}}$ denote the calculated and measured mechanical resonance frequencies, respectively.  \label{sampledimensions}}
\begin{center}
\begin{tabular}{|c|c|c|c|c|c|c|c|}
\hline
sample &$l$($\mu$m) &$t_x$(nm) &$t_y$(nm) &$x_0$(nm)&$f_0^{\m{th}}$(MHz) &$f_0^{\m{exp}}$(MHz) \\ \hline
A&4.2 &125 &141 &24 &65 &67.3 \\ \hline
B&3.0 &147 &141 &29 &114 &101.3  \\ \hline
C&5.4 &161 &143 &18 & 46 &50.8 \\\hline
D&5.3 &187 &143 &21 &52 &52.9  \\ \hline
E&9.3 &309 &155 &20 &27 &27.2  \\ \hline
F&9.3 &290 &155 &14 &27 &26.7  \\ \hline
\end{tabular}
\end{center}
\end{table}

An interesting issue in the fabrication process is that of possible side-effects of FIB milling on the sidewalls of the slit. These include Ga implantation, amorphization, heating, and generation of defects. Investigations of these artifacts have mainly concentrated on substrate materials and semiconductors~\cite{McCaffrey2001,Wang2005b}. However, some general conclusions can be made on the basis of our process parameters. The thickness of the sidewall layer with damage is on the order of 10-30~nm for the employed beam energy of 30~kV \cite{McCaffrey2001,Wang2005b,Kiener2007}. Since this is not negligible in comparison to the width of our beams, both mechanical as well as electrical properties may be altered by the FIB-induced damage.

The damage is known to lead to stress enhancement in the milled structures, due to either formation of solid solution, or Taylor or precipitation hardening~\cite{Kiener2007,Bei2007,Samayoa2008}. Ion doses required to cut the film led to bent beams, an indication of compressive strain induced in the surface exposed to milling. Stress effects were negligible for beams of relatively low aspect ratio $l/t_x \lesssim 30$, confirmed by the uniform structure in Fig.~\ref{fig:NEMS}. We use differential thermal contraction to estimate the additional stress on cooling, $T_0 \sim E t_x t_y \frac{\Delta l}{l}$ with $\frac{\Delta l}{l} \simeq 4 \cdot 10^{-3}$. Together with $E=70$ GPa and an initial stress of 0.15 GPa, we obtain resonant frequencies within 10\% of those listed in Table~\ref{sampledimensions}.

We first detected and characterized the flexural vibrations by electric actuation through the gate capacitance, see Ref.~\cite{Sillanpaa2009} as in Fig.~\ref{Readout}. In order to maximize the coupling, the stray capacitance $C$ in parallel to the beam should be minimized. For the best setup (sample E), we used an on-chip spiral coil inductor with 15 turns and 3 $\mu$m linewidth. The stray capacitance was further reduced by removal of a substantial part of the underlying substrate by the Si etch. The values $C \simeq 30$ fF, $L = 14.5$ nH were obtained from electromagnetic simulation of the structure, and which agree with the measured $f_{\m{LC}} = 7.64$ GHz. For the other samples (apart from E), we used off-chip inductors with $C \sim 0.3$ pF, and $L \sim 3 - 10$ nH. The cavity linewidth is adjusted by the tuning capacitor $C_T \sim 1-5$ pF such that $\Delta f_{\m{LC}} \lesssim f_0$.

The force driving the beam is given as $F(t) = \frac{1}{2} V_g^2(t) \frac{dC_g}{dx}$. The mechanical motion gives rise to a time-varying capacitance $C_R(t) = C_{g0} + \LL( \frac{\partial C_g}{\partial x}\RR) x$. The variation of the total capacitance of the $LC$ resonator leads to its frequency modulation, which creates sidebands at $f_{\pm} \equiv f_{\m{LC}} \pm f_0$ in a microwave reflection measurement. The blue sideband voltage $V_+$ was directly measured by a spectrum analyzer as shown in Fig.~\ref{Readout}(a). As a first amplification stage, we employed preamplifiers with a noise temperature $T_N$ varying between $\sim 3...300$ K. Two circulators in cascade were employed to reduce preamplifier backaction noise. The low-temperature measurements were performed in a dilution refrigerator that also provides an ultra high vacuum environment.

Fig.~\ref{Readout}(b) displays the driven response for a 5-$\mu$m-long Al beam with a gap of 21 nm (sample D in Table I). The peaks were measured by sweeping $f_g$ and stepping $V_1$ by 10 dB between successive spectra. We observe a linearly increasing Lorentzian response, which displays a Duffing-type instability (stiffening spring) at deflections on the order of 2 nm as expected for a doubly clamped beam resonator.

In Fig.~\ref{Readout}(c) we display the measured side band resonance position as a function of dc bias $V_{\m{dc}}$. The concave parabolic shape with gate voltage is characteristic of resonators with small tension. Since the dc voltage dependence becomes very sensitive to the vacuum gap $x_0$ at small separations, this dependence can be used for verification of the gap size. Indeed, we obtain the best fit for a gap which is close to that seen in an SEM image (18 nm), whereas larger values are clearly excluded. The solid curves display the calculated dependence modeling the capacitance $C_g$ of two parallel beams with circular cross section $C_g = \pi \epsilon_0 l \LL\{ \ln \LL[ x_0/t_y + 1 + \sqrt{\LL(x_0/t_y\RR)^2  + 2 x_0/t_y} \RR] \RR\}^{-1}$. This estimate gives, for instance, $C_g \sim 0.5$ fF for sample E. We also infer that the FIB processed regions remain metallic.

Attempts to cool mechanical systems down to the quantum ground state have usually taken place at relatively low frequencies, using active feedback cooling, or by passive sideband cooling~\cite{Karrei,Regal2008,Schwab10}. Using the latter method, occupation quanta number $\sim 4$ were recently obtained ~\cite{Schwab10}. At higher frequencies, electrical cooling is not necessary to achieve low quantum occupation, as demonstrated by the Cleland group~\cite{ClelandMartinis}. For our resonators around 50 MHz, relatively low occupancies are feasible already without sideband cooling.

\begin{figure}
  \includegraphics[width=8cm]{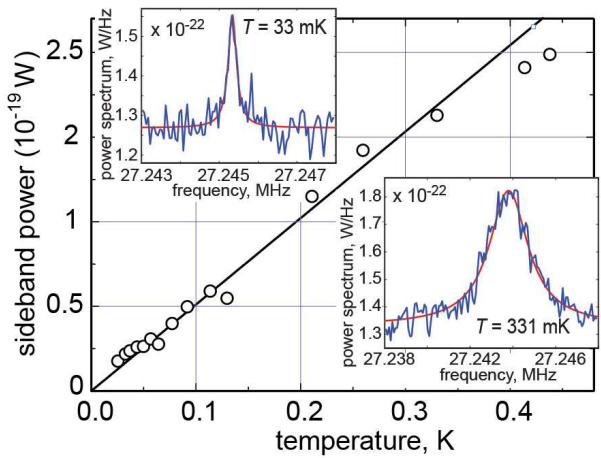}
  \caption{The total energy of the thermomechanical vibration peak measured for sample E ($f_0 = 27$ MHz) as a function of the cryostat temperature. Linear behavior is observed down to the base temperature of 26 mK, where we obtain a low number of quanta $n \sim 20$ without electrical cooling via the cavity. The insets show the thermal vibration spectra measured at two different temperatures. The red lines are fits (see text). The width of the peaks changes due to the strong temperature dependence of the mechanical quality factor.}\label{fig:Nquanta}
\end{figure}

We measured the thermomechanical vibrations at several different temperatures for the $9 \, \mu$m long beam, sample E. The temperature was simultaneously monitored by a primary Coulomb blockade thermometer. The noise floor is set by the cold amplifier noise, $T_N \sim 2.5$ K.

We fitted a Lorentzian (representing the thermal vibration peak), plus the noise floor to the measured sideband power, see insets of Fig~\ref{fig:Nquanta}. The area $A$ under the peak, given by the parameters of the fitted Lorentzian, is directly proportional to the energy of the vibration degree of freedom, but requires knowledge of the system gain which is difficult to deduce accurately. At higher temperatures, we assume that the beam is thermalized and calibrate the area of the peak against $k_B T = A = n \hbar \omega_0$. We obtained linear behavior, see Fig~\ref{fig:Nquanta}, of $A$ against $T$ down to the base temperature of 26 mK, and hence infer a low number of quanta $\sim 20$ at 26 mK.


The coupling energy $g$ between the resonator and cavity can be estimated using the parameters of the cavity, and $d C_g/dx \simeq 10$ nF/m based on the dimensions of the beam and the gap. The model is verified by the dc voltage dependence, see Fig.~\ref{Readout}. We obtain an order of magnitude larger coupling $g \sim 2 \pi \, \times 1$ MHz/nm than previously demonstrated~\cite{Regal2008,Schwab10,Teufel}.

A high mechanical quality factor $Q_m$ is instrumental for various detector applications, as well as side band cooling near the quantum ground state, and to observe phenomena in that regime. The losses are often dominated by defects which influence the mechanical and thermodynamic properties of amorphous materials at low temperatures~\cite{Phillips1987}, described using two level tunneling states (TLS). Typically, the quality factor decreases with the physical size of the resonator, an indication that surface losses pose a bottleneck~\cite{Ekinci2005}. A suitable figure of merit to take into account both frequency and $Q$ value is their product, $f_0 Q_m$. For sample A, we obtain, with the lowest mode at 67 MHz and $Q_m = 2.7 \times 10^5$ at $T = 25$ mK, $f_0 Q_m \simeq 2\times 10^{13}$ Hz.

Interestingly, we observed that the $Q$ value seemed to be improved by the FIB milling. A test sample fabricated using a single e-beam lithography step, having $l=2.5 \, \mu$m, $f_0 = 105$ MHz, displayed an order of magnitude lower $Q_m = 25 \times 10^3$ at $T = 25$ mK. The improvement of $Q$ can be due to modification of the properties of microscopic defects. A more complicated scenario could be envisioned owing to the compressive strain generated by the FIB cutting at the edges of the beam. This would cause the edges to have less strain than the rest of the beam. Therefore the stress due to flexural motion, which is usually concentrated near the beam edge would be more uniformly distributed over the beam. The coupling of phonons to defects depends on the strength of the local strain field, and thus is reduced.

\begin{figure}
  \includegraphics[width=9cm]{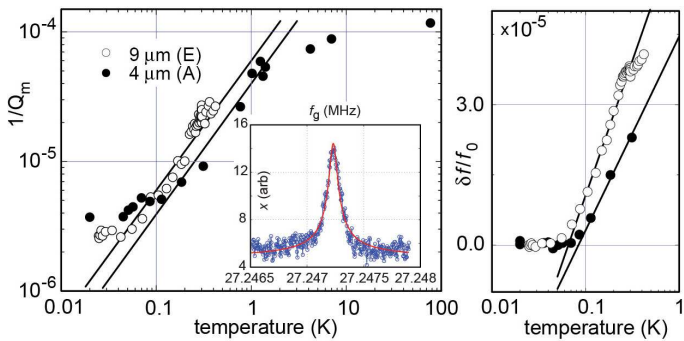}
  \caption{Measured temperature dependence of the inverse mechanical quality factor (samples A and E). The solid line indicates $\propto T$ dependence. The inset is an example of the driven response (sample E), at $T = 25$ mK, fitted to a Lorenzian which yields $Q_m \simeq 3.64 \times 10^5$. The right frame shows the corresponding change in the mechanical frequency, together with a logarithmic fits.}\label{fig:QvsTemp}
\end{figure}

We also investigated the temperature dependence of the performance of some of the samples. We used the driven response, as in Fig.~\ref{Readout}(b), or a different approach which uses two microwave frequencies $f_{\m{LC}}$ and $f_{2} $ such that $f_{2}-f_{\m{LC}} \sim f_0$~\cite{viikari}. Both methods gave similar results. In the latter case, the low-frequency force driving the mechanics is now $\propto V_{\m{LC}} V_2$, where $V_{\m{LC}}$ and $V_2$ are the microwave voltages. $f_{2}$ was kept constant, while reflection at $f_{\m{LC}}$ was swept with a vector network analyzer. We find a strong decrease in the dissipation $Q_m^{-1}$ with decreasing temperature in the range 0.1 - 2 K, see Fig.~\ref{fig:QvsTemp}. Our data follow approximately $Q_m^{-1} \propto T$ in this range. This behavior is qualitatively similar to that observed in latest experiments measured in a considerable magnetic field. Much weaker $T$-dependence was obtained in amorphous gold resonators~\cite{Venkatesan2009}. Hoehne \emph{et al.}~\cite{Pashkin10} obtained recently a similar linear temperature dependence in lithographically patterned Al resonators. Therefore, we conclude that while the dissipation is decreased due to FIB processing, its source likely stays unchanged. Notice that the saturation below 0.1 K is not due to thermal decoupling, since we observe a consistent linear dependence of the thermal vibration energy down to 26 mK, see Fig.~\ref{fig:Nquanta}.

The mechanical resonance frequency $f_0$ was also found to depend on $T$, although rather weakly: at $V_{\m{dc}}=0.2$ V, $\delta f/f_0 \simeq 2 \cdot 10^{-5} \ln \LL( T/T_{0}\RR)$ while at $V_\m{dc}=0.8$ V, we find $\delta f/f_0 \simeq 4 \cdot 10^{-5} \ln \LL( T/T_{0}\RR) $. Since $\frac{\partial f_0}{\partial V_{\m{dc}}}$ grows with $V_{\m{dc}}$, part of the frequency shift can be attributed to changes in the dielectric constant of the substrate material.

The logarithmic temperature dependence of $\delta f/f_0$ and the rolloff in $1/Q_m$ observed here (Fig.~\ref{fig:QvsTemp}) are qualitatively consistent to those predicted by TLS and explained by the tunneling model~\cite{Phillips1987}. In this model, the former is due to the resonant interaction between drive phonons and TLS. The rolloff in $1/Q_m$ is observed at the temperature $T_{\m{CO}} \sim 1$ K at which the relaxation rate of the TLS ($\tau^{-1}$) equals the mechanical frequency, $\omega_0 \tau = 1$. This is consistent with that observed in Ref.~\onlinecite{F10} since $T_{\m{CO}}$ varies as $\omega^{1/3}$.

The temperature dependence of $1/Q_m$ for $T < T_{\m{CO}}$ depends on the dominant relaxation mechanism (electron-driven or phonon-driven) and size effects. Electron-driven relaxation produces a linear temperature dependence of $1/Q_m$ for $T < T_{\m{CO}}$. However, here, in the superconducting state, one expects that phonon-driven relaxation should dominate~\cite{Fulde}. Yet the $T^3$ dependence of $1/Q_m$ expected for $T \ll T_{\m{CO}}$ due to relaxation driven by 3D phonon modes~\cite{Fefferman2008} is not developed, perhaps because other dissipation mechanisms become important at the lowest temperatures.

The dominant thermal phonon wavelength in Al, $\sim 37$ nm$\,$K/T becomes comparable to the widths and thicknesses of the beams at 100-300 mK~\cite{Klitsner}. In this limit, the dominant modes are no longer 3D phonon modes~\cite{Landau59} and can contribute a linear temperature dependence \cite{Seoanez2008,Pashkin10}. This mechanism thus could dominate relaxational damping in our high-frequency resonators. At the lowest temperatures, the resonant contribution to $1/Q_m \propto \tanh \LL(\hbar \omega/2 k_B T \RR)$ may contribute to the measured saturation.

The large capacitive coupling we have achieved between a nanoresonator and an $LC$ circuit can be useful for various sensitive detectors. The responsivity of the signal voltage to a mass change is $d V/d m \sim V Q_m/m$. Four our sample (E) set-up at $T=4$ K as in Fig.~\ref{Readout}, we obtain $s_m \sim 10^{-26}$ kg/$\sqrt{\m{Hz}} \sim 10$ AMU/$\sqrt{\m{Hz}}$, which would improve the state of the art~\cite{massdetect} of mass detection by an order of magnitude. For studies near the quantum limit, the strong coupling achieved via FIB milling is a critical asset since it enables $Q$-enhanced high-frequency resonators. Indeed, for short beams $1-2 \, \mu$m in length, we have obtained vacuum gaps approaching 5 nm. These findings substantially improve, for example, the efficiency of electrical cooling via the $LC$ circuit which is inversely proportional to the fourth power of the gap.

\textbf{Acknowledgment.}
We thank T. Heikkil\"a and H. Sepp\"a for fruitful discussions. This work was supported by the Academy of Finland, and jointly by the NSF under DMR-0908634 (Materials World Network) and the Academy of Finland, the ERC contracts FP7-240387, EU FP6-IST-021285-2, and the NANOSYSTEMS project with Nokia Research Center.


%


\begin{thebibliography}{99}%
\bibitem{Clelandbook}Cleland, A.; \emph{Foundations of Nanomechanics} (Springer, New York, 2003).
\bibitem{Ekinci2005}Ekinci, K. L.; Roukes, M. L. \emph{Rev. Sci. Instrum.} \textbf{2005}, 76, 061101.
\bibitem{SchwabQ}Schwab, K. C.; Roukes, M. L. \emph{Phys. Today} \textbf{2005}, 58, 36-42.
\bibitem{LaHaye2004} LaHaye, M. D.; Buu, O.; Camarota, B.; Schwab, K. C. \emph{Science} \textbf{2004}, 304, 74-77.
\bibitem{Karrei} Metzger, C.; Karrai, K. . \emph{Nature} \textbf{2004}, 432, 1002-1005.
\bibitem{Gigan} Gigan, S. \emph{et al.}. \emph{Nature} \textbf{2006}, 444, 67-70.
\bibitem{Kippenberg} Schliesser, A.; Del'Haye, P.; Nooshi, N.; Vahala, K. J.; Kippenberg, T. J. \emph{Phys. Rev. Lett.} \textbf{2006}, 97, 243905.
\bibitem{Knobel} Knobel, R. G.; Cleland, A. N. \emph{Nature} \textbf{2003}, 424, 291-293.
\bibitem{Truitt2007}Truitt, P. A.; Hertzberg, J. B.; Huang, C. C.; Ekinci, K. L.; Schwab, K. C. \emph{Nano Lett.} \textbf{2007}, 7, 120-126.
\bibitem{Regal2008} Regal, J. D.; Teufel, C. A.; Lehnert, K. W. \emph{Nature Physics} \textbf{2008}, 4, 555-560.
\bibitem{Sillanpaa2009}Sillanp\"a\"a, M. A.; Sarkar, J.; Sulkko, J.; Muhonen, J.; Hakonen, P. J. \emph{Appl. Phys. Lett.} \textbf{2009}, 95, 011909-011911.
\bibitem{Schwab10} Rocheleau, T.; Ndukum, T.; Macklin, C.; Hertzberg, J. B.; Clerk, A. A.; Schwab, K. C. \emph{Nature} \textbf{2009}, 463, 72-75.
\bibitem{ClelandMartinis} O'Connell, A. D. \emph{et al.}. \emph{Nature} \textbf{2010}, 464, 697-703.
\bibitem{massdetect} Jensen, K.; Kim, K.; Zettl, A. \emph{Nature Nanotechnol.} \textbf{2008}, 3, 533-537.
\bibitem{massdetectB} Lassagne, B.; Garcia-Sanchez, D.; Aguasca, A.; Bachtold, A. \emph{Nano Lett.} \textbf{2008}, 8, 3735-3738.
\bibitem{Li2008} Li, T. F.; Pashkin, Y. A.; Astafiev, O.; Nakamura, Y.; Tsai, J. S.; and Im, H.; \emph{Appl. Phys. Lett.} \textbf{2008}, 92, 043112-043114.
\bibitem{McCaffrey2001} McCaffrey, J.~P.; Phaneuf, M.~W.; Madsen, L.~D. \emph{Ultramicroscopy} \textbf{2001}, 87, 97-104.
\bibitem{Wang2005b}Wang, Z.; Kato, T.; Hirayama, T.; Kato, N.; Sasaki, K.; Saka, H. \emph{Appl. Surface Science} \textbf{2005}, 241, 80-86.
\bibitem{Kiener2007} Kiener, D.; Motz, C.; Rester, M.; Jenko, M.; Dehm, G. \emph{Materials Science and Engineering:} A \textbf{2007}, 459, 262-272.
\bibitem{Bei2007} Bei, H.; Shim, S.; Miller, M.~K.; Pharr, G.~M.; George, E.~P. \emph{Appl. Phys. Lett.} \textbf{2007}, 91, 111915-111917.
\bibitem{Samayoa2008} Samayoa, M.~J.; Haque, M.~A.; Cohen, P.~H. \emph{J. Micromechanics and Microengineering} \textbf{2008}, 18, 95005-95011.
\bibitem{Teufel}High coupling energies were also recently demonstrated in a different geometry, J. Teufel, personal communication (2010).
\bibitem{Phillips1987} Phillips, W. A. Rep. Progr. Phys.; \textbf{1987}, 50, 1657-1708.
\bibitem{Venkatesan2009} Venkatesan, A.; Lulla, K. J.; Patton, M. J.; Armour, A. D.; Mellor, C. J.; Owers-Bradley, J. R. \emph{Phys. Rev. B} \textbf{2010}, 81, 073410.
\bibitem{Pashkin10}Hoehne, F.; Pashkin, Yu. A.; Astafiev, O.; Faoro, L.; Ioffe, L. B.; Nakamura, Y.; Tsai, J. S. \emph{Phys. Rev. B} \textbf{2010}, 81, 184112.
\bibitem{viikari}Viikari V.; Sepp\"a, H. \emph{IEEE Sens.~J.} \textbf{2009}, 9, 1918 - 1923.
\bibitem{F10} Fefferman, A.; Pohl, R.~O.; Parpia, J.~M. submitted to \emph{Phys. Rev. B} \textbf{2010}.
\bibitem{Fulde} Black, J. L.; Fulde, P. \emph{Phys. Rev. Lett.} \textbf{1979}, 43, 453-456.
\bibitem{Fefferman2008} Fefferman, A. D.; Pohl, R. O.; Zehnder, A. T.; Parpia, J. M. \emph{Phys. Rev. Lett.} \textbf{2008}, 100, 195501.
\bibitem{Seoanez2008} Seoa'nez, C.; Guinea, F.; Castro Neto, A. H. \emph{Phys. Rev. B} \textbf{2008}, 77, 125107.
\bibitem{Klitsner} Klitsner, T.; Pohl, R. \emph{Phys. Rev. B} \textbf{1987}, 36, 6551-6565.
\bibitem{Landau59} Landau, L. D.; Lifshitz, E.~M.; \emph{Theory of Elasticity} (Pergamon, London, 1959).
\end{thebibliography}
\end{document}